\documentclass[pra,aps,showpacs,twocolumn]{revtex4}
\usepackage{amssymb}
\usepackage{graphicx}
\usepackage{latexsym}
\usepackage{amsmath}
\usepackage[dvipdfm]{hyperref}

\begin{document}

\title{coherence measures based on sandwiched R\'{e}nyi relative entropy}
\author{Jianwei Xu}
\email{xxujianwei@nwafu.edu.cn}
\affiliation{College of Science, Northwest A\&F University, Yangling, Shaanxi 712100,
China}
\date{\today }

\begin{abstract}
Coherence is a fundamental ingredient for quantum physics and a key resource
for quantum information theory. Baumgratz, Cramer, and Plenio established a
rigorous framework (BCP framework) for quantifying coherence [T. Baumgratz,
M. Cramer, and M. B. Plenio, Phys. Rev. Lett. \textbf{113}, 140401 (2014)].
In this paper, under the BCP framework we provide two classes of coherence
measures based on the sandwiched R\'{e}nyi relative entropy. We also prove that we can not get new coherence measures $f(C(\cdot))$
by a function $f$ acting on a given coherence measure $C$, except the case of qubit states.
\end{abstract}

\pacs{03.65.Ud, 03.67.Mn, 03.65.Aa}
\maketitle

\bigskip

\section{Introduction}

Quantum coherence is one of the most fundamental features of quantum
physics. Recently, Baumgratz, Cramer, and Plenio established a rigorous
framework (BCP framework) for quantifying coherence \cite{BCP2014}. The BCP
framework has been widely accepted and triggered rapidly growing research
for quantifying coherence (recent reviews see \cite{SAP2017,HHPZF2017}).

To construct a coherence measure, we first define the incoherent states and
incoherent operations as in \cite{BCP2014}. For a fixed orthonormal basis $%
\{|j\rangle \}_{j=1}^{d}$ of the $d$-dimensional Hilbert space $H,$ a
quantum state $\sigma $ on $H$ is called incoherent with respect to $%
\{|j\rangle \}_{j=1}^{d}$ if $\sigma $ is diagonal when expressed in $%
\{|j\rangle \}_{j=1}^{d}.$ We denote the set of all incoherent states by $%
\mathcal{I},$ and the set of density operators by $\mathcal{D}.$ A quantum
operation $\Phi $, or called a CPTP (completely positive trace preserving)
map, can be expressed by a set of Kraus operators $\{K_{n}\}_{n}$ satisfying
$\sum_{n}K_{n}^{\dag }K_{n}=I,$ where $I$ being the identity operator on $H$%
, and operate a state $\rho $ as $\Phi (\rho )=\sum_{n}K_{n}\rho K_{n}^{\dag
}.$ A quantum operation $\Phi $ is called an incoherent operation (ICPTP) $%
\Phi _{I}$, if it admits a set of Kraus operators $\{K_{n}\}_{n}$ and $%
K_{n}\sigma K_{n}^{\dag }$ is diagonal for any $n$ and any incoherent state $%
\sigma .$ Notice that the definitions of incoherent state, incoherent
operation, and also the coherence measure, are all depend on the fixed
orthonormal basis $\{|j\rangle \}_{j=1}^{d}$, we call this basis the
reference basis.

The BCP framework consists of the following postulates (C1-C4) that any
quantifier of coherence $C$ should fulfill. \newline
(C1) Non-negativity:
\begin{eqnarray}
C(\rho )\geq 0, \ \ C(\rho )=0 \Leftrightarrow \rho \in \mathcal{I}.
\end{eqnarray}
(C2) Monotonicity: $C$ does not increase under the operation of any
incoherent operation $\Phi _{I},$
\begin{eqnarray}
C[\Phi _{I}(\rho )]\leq C(\rho ).
\end{eqnarray}
(C3) Strong monotonicity: for any incoherent operation $\Phi
_{I}=\{K_{n}\}_{n},$
\begin{eqnarray}
\sum_{n}tr(K_{n}\rho K_{n}^{\dag })C[\frac{K_{n}\rho K_{n}^{\dag }}{%
tr(K_{n}\rho K_{n}^{\dag })}]\leq C(\rho ).
\end{eqnarray}
(C4) Convexity: $C$ is a convex function of the state, i.e.,
\begin{eqnarray}
\sum_{n}p_{n}C(\rho _{n})\geq C(\sum_{n}p_{n}\rho _{n}),
\end{eqnarray}
where $p_{n}>0,$ $\sum_{n}p_{n}=1,$ $\rho _{1},$ $\rho _{2}$ $\in \mathcal{D}
$.

We call a quantifier $C$ satisfying (C1-C4) together a coherence measure.
Note that C3+C4 implies C2 \cite{BCP2014}.

Yu, Zhang, Xu, and Tong proposed condition (C5), and showed that (C1-C4) is
equivalent to (C1+C2+C5) \cite{Tong2016}. \newline
(C5) additivity on block-diagonal states:
\begin{eqnarray}
C(p_{1}\rho _{1}\oplus p_{2}\rho _{2})=p_{1}C(\rho _{1})+p_{2}C(\rho _{2}),
\end{eqnarray}
where $p_{1}>0,p_{2}>0,$ $p_{1}+p_{2}=1,\rho _{1},$ $\rho _{2}$ $\in
\mathcal{D}$.\bigskip

BCP framework draws strong attention and discussion, but it is not the
unique framework for quantifying coherence, and other potential candidates
have been investigated (see  \cite{Aberg2006,Chitambar and Gour
2016a,Chitambar and Gour 2016b,Chitambar and Gour 2017,Marvian and Spekkens
2016,Winter and Yang 2016,Yadin2016,de Vicente and Streltsov 2017,Lloyd2017}
etc).

Thus far, some coherence measures have been found out for different
applications and backgrounds, such as relative entropy of coherence \cite%
{BCP2014}, the $l_{1}$ norm of coherence \cite{BCP2014}, geometric coherence
\cite{Streltsov2015}, modified trace norm of coherence \cite{Tong2016},
robustness of coherence \cite{Adesso2016-1,Adesso2016-2}, coherence measure
via quantum skew information \cite{Yu2017}, coherence measures based on
Tsallis relative entropy \cite{Rastegin2016,Yu2018,Kollas2018}, coherence
weight \cite{Bu2018}. For a coherence measure defined only for all pure
states, it can be extended to mixed states via the convex roof construction
\cite{Yuan et al. 2015, Aberg2006, Winter and Yang 2016, Qi2016}. Also, a
coherence measure defined on all pure states is determined by its
majorization property on the modular square of coefficients of pure states
\cite{Du2015,Du2017,Chitambar and Gour 2016a,Chitambar and Gour
2016b,Chitambar and Gour 2017, Zhu2017}. Although the convex roof
construction and majorization on pure states together provide a powerful way
to construct coherence measures, the coherence measures obtained in such way
generally speaking are only in the form of optimization and hard to get the
analytical expressions \cite{SAP2017}.

In this paper, we provide two classes of coherence measures based on the
sandwiched R\'{e}nyi relative entropy in Section II, the geometric coherence
\cite{Streltsov2015} is a special case of them. Also in Section III, we
discuss whether or not one can get a new coherence measure through a
function of a given coherence measure, and coherence measures for qubit
states.

\section{Coherence measures based on sandwiched R\'{e}nyi relative entropy}

In this section, we propose two classes of coherence measures based on the
sandwiched R\'{e}nyi relative entropy.   \newline

\textbf{Theorem 1.} For $\alpha \in \lbrack \frac{1}{2},1),\rho \in \mathcal{%
D},$
\begin{eqnarray}
C_{s1,\alpha }(\rho )=1-\max_{\sigma \in \mathcal{I}}(\{tr[(\rho ^{\frac{%
1-\alpha }{2\alpha }}\sigma \rho ^{\frac{1-\alpha }{2\alpha }})^{\alpha
}]\}^{\frac{1}{1-\alpha }}),
\end{eqnarray}
is a coherence measure.    \newline

\emph{Proof.} For $\alpha \in \lbrack \frac{1}{2},1),$ $\sigma ,\rho \in
\mathcal{D},$ the sandwiched R\'{e}nyi relative entropy was defined as \cite%
{Yang2014,Muller-Lennert2013}
\begin{eqnarray}
F_{\alpha }(\sigma ||\rho )=\frac{\ln tr[(\rho ^{\frac{1-\alpha }{2\alpha }%
}\sigma \rho ^{\frac{1-\alpha }{2\alpha }})^{\alpha }]}{\alpha -1}.
\end{eqnarray}
Note that $F_{\alpha }(\sigma ||\rho )$ can be defined for $\alpha >0$ \cite%
{Yang2014,Muller-Lennert2013}, but in Theorem 1 we only consider the case of
$\alpha \in \lbrack \frac{1}{2},1).$

It is shown that \cite{Muller-Lennert2013,Beigi2013} for $\alpha \in \lbrack
\frac{1}{2},1),$
\begin{eqnarray}
F_{\alpha }(\sigma ||\rho )\geq 0, \text{and equality iff } \sigma =\rho.
\end{eqnarray}
This is equivalent to
\begin{eqnarray}
tr[(\rho ^{\frac{1-\alpha }{2\alpha }}\sigma \rho ^{\frac{1-\alpha }{2\alpha
}})^{\alpha }\leq 1, \text{and equality iff } \sigma =\rho ,
\end{eqnarray}
and further equivalent to
\begin{eqnarray}
\{tr[(\rho ^{\frac{1-\alpha }{2\alpha }}\sigma \rho ^{\frac{1-\alpha }{%
2\alpha }})^{\alpha }]\}^{\frac{1}{1-\alpha }}\leq 1, \text{and equality iff
} \sigma =\rho .
\end{eqnarray}
This says that $C_{s1,\alpha }(\rho )$ satisfies (C1).

For $\alpha \in \lbrack \frac{1}{2},1),$ it has been shown that \cite%
{Muller-Lennert2013,Lieb2013} for $\sigma ,\rho \in \mathcal{D},$ and any
CPTP map $\Phi ,$
\begin{eqnarray}
F_{\alpha }(\Phi (\sigma )||\Phi (\rho ))\leq F_{\alpha }(\sigma ||\rho ).
\end{eqnarray}
This implies
\begin{eqnarray}
tr[(\Phi (\rho ))^{\frac{1-\alpha }{2\alpha }}\Phi (\sigma )(\Phi (\rho ))^{%
\frac{1-\alpha }{2\alpha }})^{\alpha }]\geq tr[(\rho ^{\frac{1-\alpha }{%
2\alpha }}\sigma \rho ^{\frac{1-\alpha }{2\alpha }})^{\alpha }], \\
\{tr[(\Phi (\rho ))^{\frac{1-\alpha }{2\alpha }}\Phi (\sigma )(\Phi (\rho
))^{\frac{1-\alpha }{2\alpha }})^{\alpha }]\}^{\frac{1}{1-\alpha }} \ \ \ \
\ \ \ \ \ \ \ \ \ \ \ \ \ \ \ \   \notag \\
\geq \{tr[(\rho ^{\frac{1-\alpha }{2\alpha }}\sigma \rho ^{\frac{1-\alpha }{%
2\alpha }})^{\alpha }]\}^{\frac{1}{1-\alpha }}. \ \ \ \ \ \ \ \ \ \ \ \ \ \
\ \ \ \ \ \ \ \ \ \ \ \ \ \ \ \
\end{eqnarray}
For any ICPTP map $\Phi _{I},$ there exists $\sigma ^{\ast }\in \mathcal{I}%
\mathbf{,}$ such that
\begin{eqnarray}
\max_{\sigma \in \mathcal{I}}\{tr[(\rho ^{\frac{1-\alpha }{2\alpha }}\sigma
\rho ^{\frac{1-\alpha }{2\alpha }})^{\alpha }]\}^{\frac{1}{1-\alpha }} \ \ \
\ \ \ \ \ \ \ \ \ \ \ \ \ \ \ \ \ \ \ \ \ \ \   \notag \\
=\{tr[(\rho ^{\frac{1-\alpha }{2\alpha }}\sigma ^{\ast }\rho ^{\frac{%
1-\alpha }{2\alpha }})^{\alpha }]\}^{\frac{1}{1-\alpha }} \ \ \ \ \ \ \ \ \
\ \ \ \ \ \ \ \ \ \ \ \ \ \ \ \ \ \ \ \ \   \notag \\
\leq \{tr[(\Phi _{I}(\rho ))^{\frac{1-\alpha }{2\alpha }}\Phi _{I}(\sigma
^{\ast })(\Phi _{I}(\rho ))^{\frac{1-\alpha }{2\alpha }})^{\alpha }]\}^{%
\frac{1}{1-\alpha }} \ \ \ \ \ \ \ \ \   \notag \\
\leq \max_{\sigma \in \mathcal{I}}\{tr[(\Phi _{I}(\rho ))^{\frac{1-\alpha }{%
2\alpha }}\sigma (\Phi _{I}(\rho ))^{\frac{1-\alpha }{2\alpha }})^{\alpha
}]\}^{\frac{1}{1-\alpha }}. \ \ \ \ \ \ \ \ \ \
\end{eqnarray}
This proves that $C_{s1,\alpha }(\rho )$ satisfies (C2).

Next we prove $C_{s1,\alpha }(\rho )$ safisfies $(C5).$ Suppose $\rho $ is
block-diagonal in the reference basis $\{|j\rangle \}_{j=1}^{d},$
\begin{eqnarray}
\rho =p_{1}\rho _{1}\oplus p_{2}\rho _{2},
\end{eqnarray}
with $p_{1}>0,p_{2}>0,p_{1}+p_{2}=1,\rho _{1},\rho _{2}\in \mathcal{D}.$

Let
\begin{eqnarray}
\sigma =q_{1}\sigma _{1}\oplus q_{2}\sigma _{2},
\end{eqnarray}
with $\sigma _{1}$, $\sigma _{2}$ diagonal states having the same rows
(columns) with $\rho _{1},$ $\rho _{2}$ respectively, $q_{1}\geq 0,q_{2}\geq
0,q_{1}+q_{2}=1.$

It follows that
\begin{eqnarray}
&&\max_{\sigma \in \mathcal{I}}tr[(\rho ^{\frac{1-\alpha }{2\alpha }}\sigma
\rho ^{\frac{1-\alpha }{2\alpha }})^{\alpha }]\ \ \ \ \ \ \ \ \ \ \ \ \ \ \
\ \ \ \ \ \ \ \ \   \notag \\
&=&\max_{q_{1},q_{2}}\{(p_{1}^{1-\alpha }q_{1}^{\alpha })\max_{\sigma
_{1}}tr[(\rho _{1}^{\frac{1-\alpha }{2\alpha }}\sigma _{1}\rho _{1}^{\frac{%
1-\alpha }{2\alpha }})^{\alpha }]\ \ \   \notag \\
&&+(p_{2}^{1-\alpha }q_{2}^{\alpha })\max_{\sigma _{2}}tr[(\rho _{2}^{\frac{%
1-\alpha }{2\alpha }}\sigma _{2}\rho _{2}^{\frac{1-\alpha }{2\alpha }%
})^{\alpha }]\}  \notag \\
&=&\max_{q_{1},q_{2}}\{p_{1}^{1-\alpha }q_{1}^{\alpha }t_{1}+p_{2}^{1-\alpha
}q_{2}^{\alpha }t_{2}\}\ \ \ \ \ \ \ \ \ \ \ \ \ \ \ \   \notag \\
&=&p_{1}^{1-\alpha }p_{2}^{1-\alpha }t_{1}t_{2}(p_{1}^{-1}t_{1}^{\frac{1}{%
\alpha -1}}+p_{2}^{-1}t_{2}^{\frac{1}{\alpha -1}})^{1-\alpha },\ \
\end{eqnarray}%
where we have denoted
\begin{eqnarray}
t_{1} &=&\max_{\sigma _{1}}tr[(\rho _{1}^{\frac{1-\alpha }{2\alpha }}\sigma
_{1}\rho _{1}^{\frac{1-\alpha }{2\alpha }})^{\alpha }, \\
t_{2} &=&\max_{\sigma _{2}}tr[(\rho _{2}^{\frac{1-\alpha }{2\alpha }}\sigma
_{2}\rho _{2}^{\frac{1-\alpha }{2\alpha }})^{\alpha }],
\end{eqnarray}%
and have used the H\"{o}lder inequality in Appendix A (note that $%
t_{1}>0,t_{2}>0).$

Consequently,
\begin{eqnarray}
&&\max_{\sigma \in \mathcal{I}}(\{tr[(\rho ^{\frac{1-\alpha }{2\alpha }%
}\sigma \rho ^{\frac{1-\alpha }{2\alpha }})^{\alpha }]\}^{\frac{1}{1-\alpha }%
})\ \ \ \ \ \ \ \ \ \   \notag \\
&=&\{\max_{\sigma \in \mathcal{I}}tr[(\rho ^{\frac{1-\alpha }{2\alpha }%
}\sigma \rho ^{\frac{1-\alpha }{2\alpha }})^{\alpha }]\}^{^{\frac{1}{%
1-\alpha }}}\ \ \ \ \ \ \ \ \ \ \   \notag \\
&=&p_{1}p_{2}t_{1}^{\frac{1}{1-\alpha }}t_{2}^{\frac{1}{1-\alpha }%
}(p_{1}^{-1}t_{1}^{\frac{1}{\alpha -1}}+p_{2}^{-1}t_{2}^{\frac{1}{\alpha -1}%
})\ \ \   \notag \\
&=&p_{1}t_{1}^{\frac{1}{1-\alpha }}+p_{2}t_{2}^{\frac{1}{1-\alpha }}.\ \ \ \
\ \ \ \ \ \ \ \ \ \ \ \ \ \ \ \ \ \ \ \ \ \
\end{eqnarray}%
This shows that $C_{s1,\alpha }(\rho )$ satisfies (C5). $\Box$

We remark that when $\alpha =2,C_{s1,\frac{1}{2}}(\rho )$ coresponds to the
geometric coherence \cite{Streltsov2015}.\bigskip

\emph{Example 1.} For pure state $\rho =|\psi \rangle \langle \psi |,$
\begin{equation}
\begin{aligned} C_{s1,\alpha }(|\psi \rangle )=1-\max_{j}\{|\langle j|\psi
\rangle |^{\frac{2\alpha }{1-\alpha }}\}. \end{aligned}
\end{equation}%
\emph{Proof.} Suppose $\sigma =\sum_{j}\sigma _{j}|j\rangle \langle j|$ is
an incoherent state,
\begin{eqnarray} tr[(\rho ^{\frac{1-\alpha }{2\alpha }}\sigma \rho
^{\frac{1-\alpha }{2\alpha }})^{\alpha }] \ \ \ \ \ \ \ \ \ \  \ \ \ \ \ \ \ \ \ \  \nonumber \\
=tr[(|\psi \rangle \langle \psi |\sum_{j}\sigma
_{j}|j\rangle \langle j|\psi \rangle \langle \psi |)^{\alpha }] \ \ \ \ \nonumber \\
=(\sum_{j}\sigma _{j}|\langle j|\psi \rangle |^{2})^{\alpha },
\ \ \ \ \ \ \ \ \ \  \ \ \ \ \ \ \ \ \ \  \
\end{eqnarray}
then we can get the result. \newline

\textbf{Theorem 2.} For $\alpha \in \lbrack \frac{1}{2},1)\cup (1,\infty ),$
\begin{eqnarray}
C_{s,\alpha }(\rho ) &=&\min_{\sigma \in \mathcal{I}}\frac{\{tr[(\sigma ^{%
\frac{1-\alpha }{2\alpha }}\rho \sigma ^{\frac{1-\alpha }{2\alpha }%
})^{\alpha }]\}^{\frac{1}{\alpha }}-1}{\alpha -1}, \\
\text{supp}(\rho ) &\subset &\text{supp}(\sigma )\text{ when }\alpha >1,
\notag
\end{eqnarray}%
is a coherence measure.  \newline

\emph{Proof.} Theorem 2 can be proved in the similar way of proof for
Theorem 1 with minor modification.$\Box$

Note that since the quantum fidelity
\begin{equation*}
F(\rho ,\sigma )=tr[(\sigma ^{\frac{1}{2}}\rho \sigma ^{\frac{1}{2}})^{\frac{%
1}{2}}]=tr[(\rho ^{\frac{1}{2}}\sigma \rho ^{\frac{1}{2}})^{\frac{1}{2}}],
\end{equation*}%
then $C_{s,\frac{1}{2}}(\rho )=C_{s1,\frac{1}{2}}(\rho )$ again coresponds
to the geometric coherence \cite{Streltsov2015}.

We remark that a coherence quantifier based on sandwiched R\'{e}nyi relative
entropy was also investigated in \cite{Chitambar and Gour 2016a,Shao2017},
but that quantifier is not a coherence measure in the sense that satisfying
(C1-C4), i.e., under the BCP framework.

\emph{Example 2.} For pure state $\rho =|\psi \rangle \langle \psi |,$
\begin{equation*}
C_{s,\alpha }(|\psi \rangle )=\frac{(\sum_{j}|\langle \psi |j\rangle |^{%
\frac{2\alpha }{2\alpha -1}})^{\frac{2\alpha -1}{\alpha }}-1}{\alpha -1}.
\end{equation*}

\emph{Proof.} Suppose $\sigma =\sum_{j}\sigma _{j}|j\rangle \langle j|$ is
an incoherent state,
\begin{eqnarray}
&&tr[(\sigma ^{\frac{1-\alpha }{2\alpha }}\rho \sigma ^{\frac{1-\alpha }{%
2\alpha }})^{\alpha }]\ \ \ \ \ \ \   \notag \\
&=&tr[(\sigma ^{\frac{1-\alpha }{2\alpha }}|\psi \rangle \langle \psi
|\sigma ^{\frac{1-\alpha }{2\alpha }})^{\alpha }]  \notag \\
&=&(\langle \psi |\sigma ^{\frac{1-\alpha }{\alpha }}|\psi \rangle )^{\alpha
}\ \ \ \ \ \ \ \ \ \ \   \notag \\
&=&(\sum_{j}\sigma _{j}^{\frac{1-\alpha }{\alpha }}|\langle \psi |j\rangle
|^{2})^{\alpha }.\ \ \ \
\end{eqnarray}%
Using the H\"{o}lder inequality (Appendix A), we then get the result.

Examples 1 and 2 show that $C_{s,\alpha }$ and $C_{s1,\beta }$ are not
equivalent even $\alpha =\beta .$

\section{Linearization theorem and coherence measures for qubit states}

One might ask that for given coherence measure $C,$ whether or not there
exists a function $f$ such that $f[C(\rho )]$ still is a coherence measure. The
answer of this question is essentially no. We have the following theorem.  \newline

\textbf{Theorem 3.} (Linearization Theorem) Given a coherence measure $C$ defined
on $d$-dimensional quantum states with $d>2$, the function $f:[0,\infty
)\longrightarrow \lbrack 0,\infty ),$ makes $f(C(\rho ))$ also a
coherence measure, if and only if there must exists $\lambda >0$, such that $%
f(x)=\lambda x.$    \newline

\emph{Proof.} Since $C(\rho )$ and $f(C(\rho ))$ are all coherence measures, then
(C1) leads to
\begin{eqnarray}
f(x)\geq 0 \ \ \text{and} \ f(0)=0 \ \ \text{iff} \ \ x=0.
\end{eqnarray}

Suppose $d>2,$ for the state
\begin{eqnarray}
\rho =p_{1}\rho _{1}\oplus p_{2}\rho _{2},
\end{eqnarray}
with $p_{1}>0, p_{2}>0, p_{1}+p_{2}=1,\rho _{1},\rho _{2}\in \mathcal{D},$ dim$%
\rho _{1}\geq 1, $ dim$\rho _{2}\geq 1,$  dim$\rho _{1}+dim\rho _{2}=d,$  since $%
C(\rho )$ and $f(C(\rho ))$ are all coherence measures, then (C5) leads to
\begin{eqnarray}
f[C(p_{1}\rho _{1}\oplus p_{2}\rho _{2})]  \ \ \ \ \ \ \ \ \     \notag \\
=f[p_{1}C(\rho _{1})+p_{2}C(\rho _{2})]  \ \ \ \    \notag \\
=p_{1}f[C(\rho _{1})]+p_{2}f[C(\rho _{2})].
\end{eqnarray}
Without loss of generality, suppose dim$\rho _{2}\geq 2,$  then there exist $%
\rho _{1},$ $\rho _{2}$ such that $C(\rho _{1})=0,C(\rho _{2})=\mu >0.$ The
above equation yields
\begin{eqnarray}
f(p_{2}\mu )=p_{2}f(\mu ),
\end{eqnarray}
with $f(\mu )>0.$

Let $p_{2}\mu =x,$ then
\begin{eqnarray}
f(x)=x\frac{f(\mu )}{\mu }=\lambda x,\lambda >0.
\end{eqnarray}
We then complete this proof. $\Box$

Since coherence measures $C$ and $\lambda C$ $(\lambda >0)$ have no
essential difference, then from Theorem 3, we say that, it is impossible to
get a new coherence measure $f(C(\rho ))$ by a function $f.$

The reason of assuming $d>2$ in Theorem 3 is that (C5) is trivial for $d=2.$
With this in mind, and after some algebra, we have the following Theorem.  \newline

\textbf{Theorem 4.} Given a coherence measure $C$ for qubit states, the function $%
f:[0,\infty )\longrightarrow \lbrack 0,\infty ),$ makes $f(C(\rho ))$ also a coherence measure, if and only if

(1). $f(x)\geq 0$ and $f(0)=0$ iff $x=0.$

(2). $f(x)\geq f(y)$ when $x\geq y.$    \newline

For example, for $d=2,$ the coherence of $l_{1}$ norm \cite{BCP2014} $C_{l_{1}}=2|\rho _{12}|=2|\langle 1|\rho |2\rangle |,$
then any function $f$ satisfying (1) and (2) of Theorem 4, makes $f(C_{l_{1}})$ still a
coherence measure, such as geometric coherence for $d=2$ \cite{Streltsov2015}
and coherence formation for $d=2$ \cite{Yuan et al. 2015}$.$

\section{SUMMARY}

In summary, under the BCP framework for quantifying coherence, we proposed
two classes of coherence measures based on sandwiched R\'{e}nyi relative
entropy. Our strategy to prove these coherence measures satisfying the
(C1-C4) of BCP framework is to prove they satisfy (C1+C2+C5). We also proved
that it is essentially impossible to get new coherence measures $f(C(\rho ))$
by a function $f$ acting on a given coherence measure $C,$ except the case
of qubit states.

There are many open questions for future investigations. For example, the
monotonicity of $C_{s1,\alpha },C_{s,\alpha }$ in $\alpha ,$ the ordering of
magnitude for them and other coherence measures, the operational
interpretations for them, potential applications in quantum information
processings, and also the counterparts for quantifying coherence of Gaussian states as done in \cite{Xu2016,Illuminati2016}. 

\section*{ACKNOWLEDGMENTS}

The author thanks Lin Zhang, Dong Yang, Lianhe Shao, Nikolaos K. Kollas, Sumiyoshi Abe, and Chandrashekar Radhakrishnan for helpful discussions and comments.

\section*{APPENDIX A: H\"{O}LDER INEQUALITY}

Suppose $\{a_{j}\}_{j=1}^{d},\{b_{j}\}_{j=1}^{d},$ are all positive real
numbers, then

(1). when $\alpha \in (0,1),$
\begin{eqnarray}
\sum_{j=1}^{d}a_{j}b_{j}\leq (\sum_{j=1}^{d}a_{j})^{\alpha
}(\sum_{j=1}^{d}b_{j})^{1-\alpha },
\end{eqnarray}
and equality iff $\frac{a_{j}}{b_{j}}=\frac{a_{k}}{b_{k}}$ for any $j,k;$

(2). when $\alpha >1,$
\begin{eqnarray}
\sum_{j=1}^{d}a_{j}b_{j}\geq (\sum_{j=1}^{d}a_{j})^{\alpha
}(\sum_{j=1}^{d}b_{j})^{1-\alpha },
\end{eqnarray}
and equality iff $\frac{a_{j}}{b_{j}}=\frac{a_{k}}{b_{k}}$ for any $j,k.$


\begin{thebibliography}{99}
\bibitem{BCP2014} T. Baumgratz, M. Cramer, and M. B. Plenio, Phys. Rev.
Lett. \textbf{113}, 140401 (2014).

\bibitem{SAP2017} Alexander Streltsov, Gerardo Adesso, and Martin B. Plenio,
Rev. Mod. Phys. \textbf{89}, 041003 (2017).

\bibitem{HHPZF2017} M. Hu, X. Hu, J. Wang, Y. Peng, Y. Zhang, and H. Fan,
arXiv:quant-ph/1703.01852.

\bibitem{Tong2016} X.-D. Yu, D.-J. Zhang, G. F. Xu, and D. M. Tong, Phys.
Rev. A \textbf{94}, 060302 (2016).

\bibitem{Aberg2006} J. Aberg, arXiv:quant-ph/0612146.

\bibitem{Chitambar and Gour 2016a} E. Chitambar, and G. Gour, Phys. Rev.
Lett. \textbf{117}, 030401 (2016).

\bibitem{Chitambar and Gour 2016b} E. Chitambar, and G. Gour, Phys. Rev. A
\textbf{94}, 052336 (2016).

\bibitem{Chitambar and Gour 2017} E. Chitambar, and G. Gour, Phys. Rev. A
\textbf{95}, 019902 (2017).

\bibitem{Marvian and Spekkens 2016} I. Marvian, and R. W. Spekkens, Phys.
Rev. A \textbf{94}, 052324 (2016).

\bibitem{de Vicente and Streltsov 2017} J. I. de Vicente, and A. Streltsov,
J. Phys. A \textbf{50}, 045301 (2017).

\bibitem{Lloyd2017} Z.-W. Liu, X. Hu, and S. Lloyd, Phys. Rev. Lett. \textbf{%
118}, 060502 (2017).

\bibitem{Winter and Yang 2016} A. Winter and D. Yang, Phys. Rev. Lett. \textbf{116},
120404 (2016).

\bibitem{Yadin2016} B. Yadin, J. Ma, D. Girolami, M. Gu, and V. Vedral,
Phys. Rev. X \textbf{6}, 041028 (2016).

\bibitem{Streltsov2015} A. Streltsov, U. Singh, H. S. Dhar, M. N. Bera, and
G. Adesso, Phys. Rev. Lett. \textbf{115}, 020403 (2016).

\bibitem{Adesso2016-1} C. Napoli, T. R. Bromley, M. Cianciaruso, M. Piani,
N. Johnston, and G. Adesso, Phys. Rev. Lett. \textbf{116}, 150502 (2016).

\bibitem{Adesso2016-2} Piani, M., M. Cianciaruso, T. R. Bromley, C. Napoli,
N. Johnston, and G. Adesso, Phys. Rev. A \textbf{93}, 042107 (2016).

\bibitem{Yu2017} C.-S. Yu, Phys. Rev. A \textbf{95}, 042337 (2016).

\bibitem{Rastegin2016} A. E. Rastegin, Phys. Rev. A \textbf{93}, 032136
(2016).

\bibitem{Yu2018} H. Zhao, and C-S. Yu, Sci. Rep. \textbf{8}, 299 (2018).


\bibitem{Kollas2018} N. K. Kollas, Phys. Rev. A \textbf{97}, 062344 (2018).

\bibitem{Bu2018} K. Bu, N. Anand, and U. Singh, Phys. Rev. A \textbf{97},
032342 (2018).

\bibitem{Yuan et al. 2015} X. Yuan, H. Zhou, Z. Cao, and X. Ma, Phys. Rev. A
\textbf{92}, 022124 (2015).

\bibitem{Qi2016} X. Qi, T. Gao, F. Yan, J. Phys. A: Math. Theor. \textbf{50}%
, 285301 (2017).

\bibitem{Du2015} S. Du, Z. Bai, and Y. Guo, Phys. Rev. A \textbf{91},
052120 (2015).

\bibitem{Du2017} S. Du, Z. Bai, and Y. Guo, Phys. Rev. A \textbf{95},
029901 (2017).

\bibitem{Zhu2017} H. Zhu, Z. Ma, Z. Cao, S.-M. Fei, and Vlatko Vedral, Phys.
Rev. A \textbf{96}, 032316 (2017).


\bibitem{Yang2014} M. M. Wilde, A. Winter, and D. Yang, Commun. Math. Phys.
\textbf{331}, 593 (2014).

\bibitem{Muller-Lennert2013} M. M\"{u}ller-Lennert, F. Dupuis, O. Szehr, S.
Fehr, and M. Tomamichel, J. Math. Phys. \textbf{54}, 122203 (2013).

\bibitem{Beigi2013} S. Beigi, J. Math. Phys. \textbf{54}, 122202 (2013).

\bibitem{Lieb2013} R. L. Frank and E. H. Lieb, J. Math. Phys. \textbf{54},
122201 (2013).
\bibitem{Shao2017} L.-H. Shao, Y.-M. Li, Yu Luo, and Z.-J. Xi, Commun.
Theor. Phys. \textbf{67} 631 (2017).
\bibitem{Xu2016} J. Xu, Phys. Rev. A \textbf{93}, 032111 (2016).
\bibitem{Illuminati2016} D. Buono, G. Nocerino, G. Petrillo, G. Torre, G. Zonzo, and F. Illuminati, arXiv:quant-ph/1609.00913.
\end{thebibliography}
\end{document}